\documentclass[aps,pra,reprint,nofootinbib,twocolumn,superscriptaddress,showpacs,showkeys,longbibliography]{revtex4-1}

\usepackage{eurosym}
\usepackage{amsmath,amssymb,amstext}
\usepackage{ulem}
\usepackage[usenames,dvipsnames]{color}
\usepackage{graphicx}
\usepackage{braket}
\usepackage{natbib}
\usepackage{comment}
\usepackage{dcolumn}
\usepackage[english]{babel}
\usepackage{wasysym}
\usepackage{bm}
\usepackage[colorlinks,bookmarks=false,citecolor=blue,linkcolor=red,urlcolor=blue]{hyperref}

\usepackage{appendix}
\usepackage{CJK}

\begin{document}
\title{Improved laser-distillation method for complete enantio-conversion of chiral mixtures}
\author{Chong Ye}
\affiliation{Beijing Key Laboratory of Nanophotonics and Ultrafine Optoelectronic Systems, School of Physics, Beijing Institute of Technology, 100081 Beijing, China}
\affiliation{Beijing Computational Science Research Center, Beijing 100193, China}
\author{Yu-Yuan Chen}
\affiliation{Beijing Computational Science Research Center, Beijing 100193, China}
\author{Quansheng Zhang}
\affiliation{Beijing Computational Science Research Center, Beijing 100193, China}
\author{Yong Li}\email{liyong@csrc.ac.cn}
\affiliation{Beijing Computational Science Research Center, Beijing 100193, China}
\affiliation{Synergetic Innovation Center for Quantum Effects and Applications, Hunan Normal University, Changsha 410081, China}

\begin{abstract}
  Laser-assisted enantio-conversion is an ambitious issue related to chiral molecules in the atomic, molecular, and optical physics. The theoretical laser-distillation method had been proposed to realize enantio-conversion based on a four-level double-$\Delta$ model of two degenerated chiral ground states and two achiral excited states. Here, we re-investigate and improve the laser-distillation method so that a chiral mixture can be converted to an enantiopure sample of the desired  chirality, i.e., complete enantio-conversion, which has not been discussed in the previous theoretical works of the laser-distillation method. Since the undesired chirality may be inefficient or even cause serious side effects in pharmacology, our work plays as an important improvement to the original theoretical works of the laser-assisted enantio-conversion.
\end{abstract}
\date{\today}
\maketitle
\section{Introduction}
Attributing to the breakthrough experiments devoted to enantio-discrimination~\cite{Nature.497.475,PRL.111.023008,PCCP.16.11114,ACI,JCP.142.214201,JPCL.6.196,JPCL.7.341} and enantio-specific state-transfer~\cite{PRL.118.123002,Angew.Chem.56.12512} based on the $\Delta$-type electric-dipole transitions of chiral molecules, more and more theoretical works tend to
address the related issues of chiral molecules~\cite{JCP.149.094201,KK,1501.05282,anie.201307159,PRA.97.033403,PRA.98.063401,PRL.122.173202,JCP.151.014302,PRA.100.033411,PRA.100.043403,PRA.100.043413}.
This tendency can trace back to the theoretical work in 2001~\cite{PRL.87.183002}, where the $\Delta$-type model of chiral molecules was first introduced to realize enantio-separation.
After that, a new avenue~\cite{PRL.90.033001,PRL.99.130403,PRA.77.015403,JCP.132.194315,JPB.43.185402,
PRA.84.053849,JCP.137.044313,PRA.100.043413,PRAp.13.044021,JCP.152.204305} concerning the enantio-discrimination and enantio-separation in the atomic, molecular, and optical physics was established, since molecular chirality plays significant roles in vast majority of chemical~\cite{A1}, biological~\cite{A2,A3,A4}, and pharmaceutical~\cite{A5,A7,A8,A6} processes.

The $\Delta$-type model of chiral molecules is special since the sign of the product of the three coupling strengths changes with enantiomers. With such a spirt, a four-level double-$\Delta$ model of chiral molecules~\cite{PRL.84.1669} was introduced to address a more
ambitious issue of laser-assisted enantio-conversion even before the first introduction of the $\Delta$-type model of chiral molecules~\cite{PRL.87.183002}. The laser-assisted
enantio-conversion~\cite{PRL.84.1669} is devoted to converting the undesired enantiomer in a chiral mixture to the desired enantiomer. In the four-level double-$\Delta$ model,
each of the two degenerated (left- and right-handed) chiral ground states, a symmetric achiral excite state, and an asymmetric
achiral excited state are coupled with three electromagnetic fields via electric-dipole
transitions in the $\Delta$-type structure. {Since the overall phases of the three coupling strengths in the two $\Delta$-type sub-structures differ with $\pi$}, the left-right symmetry of this four-level model is broken. Thus the initial chiral ground states with the same populations can be finally de-populated differently in the framework of the four-level double-$\Delta$ model~\cite{PRL.84.1669}. Such a symmetry breaking in the de-population offers the possibility of converting one enantiomer to another.

In the pioneer theoretical works of laser-assisted enantioconversion~\cite{PRL.84.1669,PRA.65.015401,JCP.115.5349,JPB.37.2811},
the symmetry-breaking feature of the four-level double-$\Delta$ model is used to realize
the laser-assisted enantio-conversion of a racemic sample in the scenario called ``laser distillation of
chiral enantiomers'' by repeating a pair of excitation and relaxation steps. In the excitation step,
the three electromagnetic fields are applied to construct the four-level double-$\Delta$ model.
After that, the three electromagnetic fields are turned off, the system experiences a relaxation step. The molecules in the achiral excited states equally relax back to the two degenerated chiral ground states~\cite{PRL.84.1669,PRA.65.015401,JCP.115.5349,JPB.37.2811}. Then, some molecules in the undesired chiral ground state are converted to the desired ones after applying this pair of steps to a racemic sample. By repeating the pair of steps, the molecules in the desired chirality will accumulate until the equilibrium is reached. {Since the desired chirality may be inefficient~\cite{A6} or even cause serious side effects~\cite{A7,A8} in pharmacology, it is natural to pursue the complete enantio-conversion, i.e., converting all molecules in the chiral mixture to the desired chirality, which has not been discussed in the previous theoretical works of laser-distillation method~\cite{PRL.84.1669,PRA.65.015401,JCP.115.5349,JPB.37.2811}.}

Based on the similar four-level double-$\Delta$ model~\cite{PRL.84.1669,PRA.65.015401,JCP.115.5349,JPB.37.2811}, we propose a theoretical work to improve the laser-distillation method for the complete enantio-conversion in the equilibrium by well designing the electromagnetic fields in the excitation step following our recent work~\cite{arXiv.xxx}.
In the well-designed scheme, the two chiral ground states evolve following two-level models
in separated two-state subspaces with the same effective coupling strengths but different effective
detunings. We propose coherence operation in each excitation step, which makes the desired chiral ground state evolve back to itself and in the meanwhile part of population in the undesired chiral ground state will be excited to the up-level of its corresponding two-level sub-system. In the relaxation step, the part of population excited from the undesired chiral state will equally relax to the two chiral ground states. Then, molecules of the desired chirality will accumulate by repeating the pair of steps and eventually the enantiopure sample is achieved when the equilibrium is reached. Unlike the original theoretical works of laser-distillation method~\cite{PRL.84.1669,PRA.65.015401,JCP.115.5349,JPB.37.2811}, which only offered
the results in the equilibrium with infinite repeated times, we also study more realistic cases with
finite repeated times. In order to optimize the efficiency in the realistic case, we pave way for further
designing the electromagnetic fields in two cases where the driving fields are in square pulses and non-square pulses, respectively.


\section{Four-level double-$\Delta$ model}
\begin{figure}[h]
  \centering
  \includegraphics[width=0.9\columnwidth]{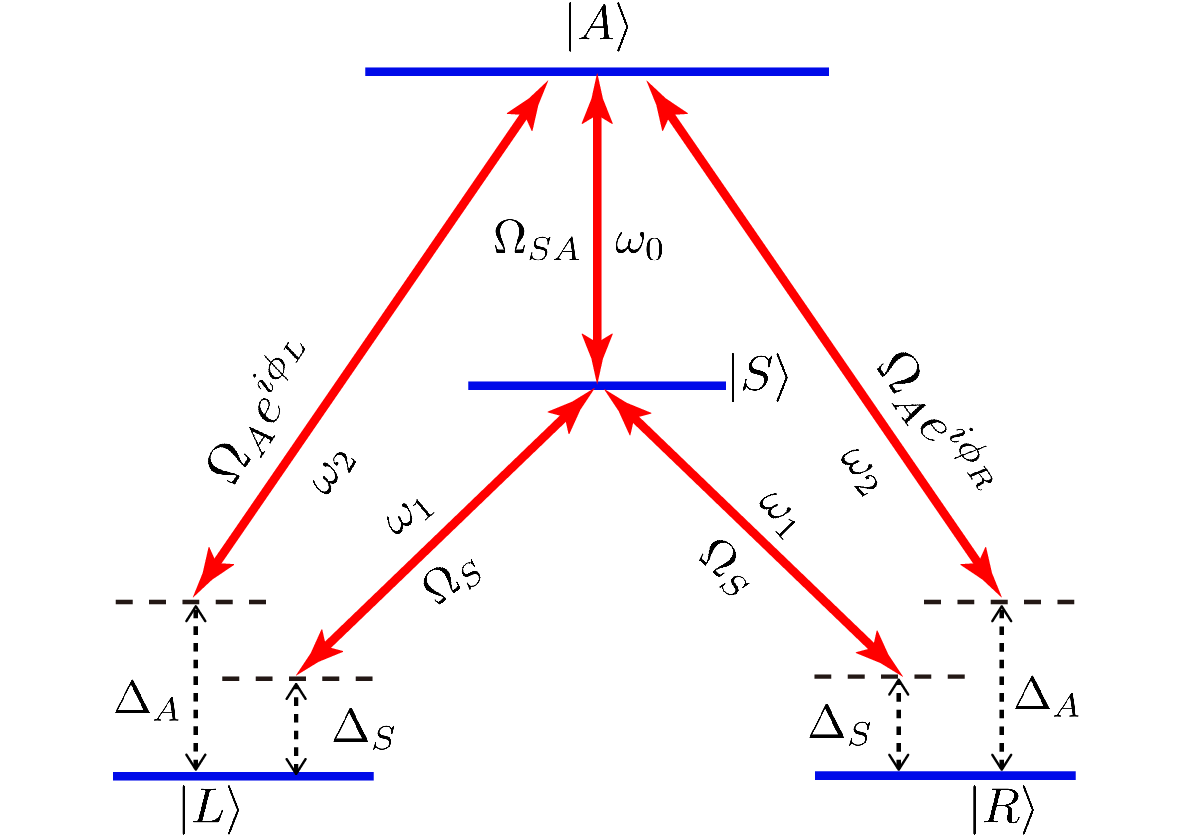}\\
  \caption{ The four-level double-$\Delta$ model of chiral molecules. $|L\rangle$ and $|R\rangle$ are, respectively, the degenerated left- and right-handed chiral ground states. $|A\rangle$ and $|S\rangle$ are, respectively, the symmetric and asymmetric achiral excited states. Three electromagnetic fields with frequencies $\omega_0$, $\omega_1$, and $\omega_2$ are applied to couple the four-level model in $\Delta$-type transitions of  $|Q\rangle\leftrightarrow|A\rangle\leftrightarrow|S\rangle\leftrightarrow|Q\rangle$ with $Q=L,R$ under three-photon resonance condition. 
  $\Delta_{A}$ and $\Delta_{S}$ are the detunings. $|A\rangle\leftrightarrow|S\rangle$ is on resonance. $\Omega_{A}$, $\Omega_{S}$, and $\Omega_{SA}$ are amplitudes of the coupling strengths corresponding to the three transitions. The overall phase for two $\Delta$-type sub-structures are $\phi_{L}$ and $\phi_{R}$ with the left-right symmetry breaking reflecting in $\phi_{R}=\phi_{L}+\pi$.}\label{Fig1}
\end{figure}
We consider the four-level  double-$\Delta$ model of chiral molecules~\cite{PRL.84.1669,PRA.65.015401,JCP.115.5349,JPB.37.2811} as shown in Fig.~\ref{Fig1}.
Transitions among each of the degenerated two chiral ground states (the left-handed one $|L\rangle$ and the right-handed one $|R\rangle$) and the two achiral excited states (the symmetric one $|S\rangle$ and the asymmetric one $|A\rangle$)
are coupled with three electromagnetic fields in the $\Delta$-type manner as
$|Q\rangle\leftrightarrow|A\rangle\leftrightarrow|S\rangle\leftrightarrow|Q\rangle$ with $Q=L,R$.
The energies of the four states are $\hbar\omega_{A}> \hbar\omega_{S}> \hbar\omega_{R}= \hbar\omega_{L}=0$.
Here, the two chiral ground states are degenerate since we have neglected the
tiny parity violating energy differences due to the fundamental weak force. For simplicity, the couplings among vibrational, rotational, and electronic states under field-free conditions are assumed negligible and the molecules are in the electronic ground state throughout~\cite{JCP.115.5349,JCP.137.044313,PRL.84.1669,PRA.65.015401,JCP.115.5349,JPB.37.2811}.
The three electromagnetic fields are under the three-photon resonance condition and the one-photon resonance condition of $|A\rangle\leftrightarrow|S\rangle$
\begin{align}\label{C1}
\omega_{2}=\omega_{1}+\omega_{0},~~\omega_{A}-\omega_{S}=\omega_{0}
\end{align}
with frequencies of the three electromagnetic fields being $\omega_0$, $\omega_1$, and $\omega_2$ as shown in Fig.~\ref{Fig1}.

In the rotating wave approximation, the Hamiltonian of the four-level  double-$\Delta$ model in the interaction picture with respect to $\hat{H}_{0}= \omega_{2}|A\rangle\langle A|+\omega_1|S\rangle\langle S|$ is given as ($\hbar=1$)
\begin{align}\label{HT}
\hat{H}=&\sum_{Q=L,R}(\Omega_{A}e^{i\phi_{Q}}|Q\rangle\langle A|+\Omega_{S}|Q\rangle\langle S|+\mathrm{H.c.})\nonumber\\
&+(\Omega_{SA}|S\rangle\langle A|+\mathrm{H.c.})+\sum_{P=A,S}\Delta_{P}|P\rangle\langle P|.
\end{align}
Here, the detunings are defined as ($\omega_{R}=\omega_{L}=0$)
\begin{align}
&\Delta_S\equiv \omega_S-\omega_1,~~\Delta_A\equiv \omega_A-\omega_2.
\end{align}
The three-photon resonance of the three electromagnetic fields and one-photon resonance of the transition $|A\rangle\leftrightarrow|S\rangle$ conditions~(\ref{C1}) yield
\begin{align}
\Delta_{A}=\Delta_{S}\equiv\Delta.
\end{align}
For simplicity and without loss of
generality, we assume $\Omega_{SA}$, $\Omega_{A}$, and $\Omega_{S}$ are positive.
Thus, the breaking of left-right symmetry in the two $\Delta$-type sub-structures~\cite{PRL.84.1669,PRA.65.015401,JCP.115.5349,JPB.37.2811} are reflected in their overall phases
\begin{align}
\phi_{R}=\phi_{L}+\pi,~~~\phi_{L}\equiv\phi.
\end{align}

\section{Effective two-level models}
Based on the four-level double-$\Delta$ model of chiral molecules, the laser-distillation
method~\cite{PRL.84.1669,PRA.65.015401,JCP.115.5349,JPB.37.2811} was introduced to realize the enantio-conversion by utilizing the chirality-dependent de-population due to the breaking of left-right symmetry in the two $\Delta$-type sub-structures. Since the two $\Delta$-type sub-structures share the transition $|A\rangle\leftrightarrow|S\rangle$, the de-population of one chiral ground state in the chiral mixture is related to the dynamics of the two chiral ground state~\cite{PRL.84.1669,PRA.65.015401,JCP.115.5349,JPB.37.2811}. In the following, we will {give} two schemes with well-designed electromagnetic fields, in which
the dynamics of each initial chiral ground state is constrained in its own two-dimensional subspace and describes by its corresponding effective two-level model in the dressed state basis. Within such simplified four-level models, we can improve the laser-distillation
method~\cite{PRL.84.1669,PRA.65.015401,JCP.115.5349,JPB.37.2811} to achieve complete enantio-conversion in equilibrium.

\begin{figure}[h]
  \centering
  \includegraphics[width=0.90\columnwidth]{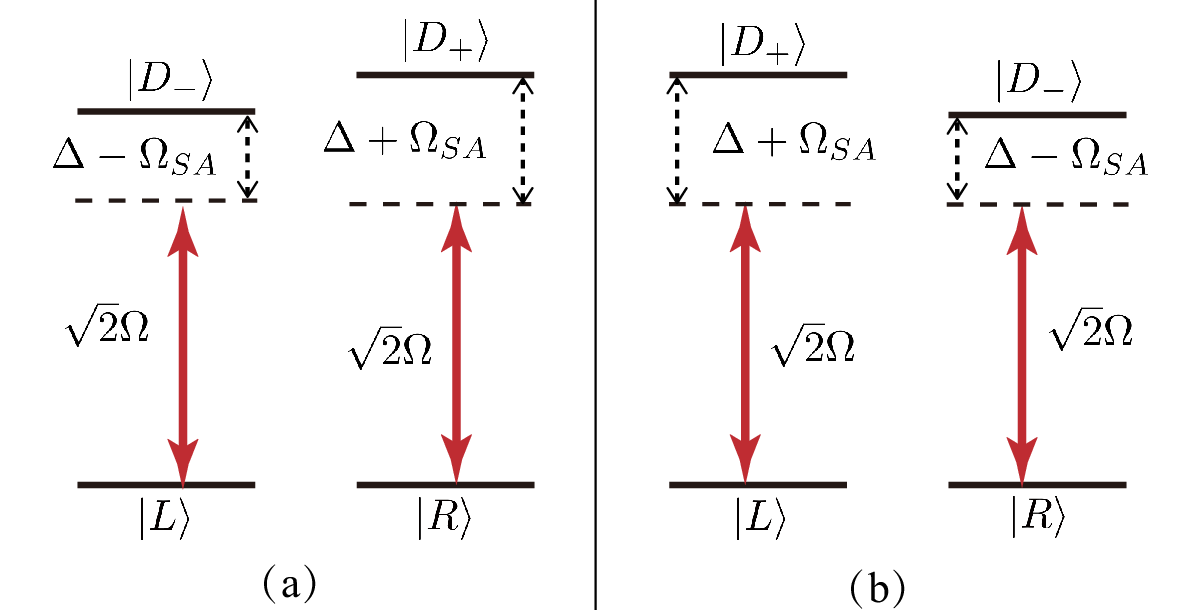}\\
  \caption{Simplified four-level model in the dressed state basis under conditions $\Delta_{A}=\Delta_{S}\equiv\Delta$ and $\Omega_{S}=\Omega_{A}\equiv\Omega$ with the well-designed overall phase
  (a) $\phi=\pi$; (b) $\phi=0$.}\label{Fig2}
\end{figure}

The strengths of the electromagnetic fields are further
tuned to ensure that
\begin{align}\label{CDI1}
\Omega_{S}=\Omega_{A}\equiv\Omega.
\end{align}
%
For the case with $\phi=\pi$, the Hamiltonian of the original four-level  double-$\Delta$ model can be rewritten in the following form
\begin{align}\label{HMR}
\hat{H}^{\mathrm{I}}=&\sqrt{2}\Omega(|L\rangle\langle D_{-}|+|R\rangle\langle D_{+}|+\mathrm{H.c.})\nonumber\\
&+\sum_{\lambda=\pm}\Delta_{\lambda}|D_{\lambda}\rangle\langle D_{\lambda}|,
\end{align}
where the dressed states $|D_{\pm}\rangle$ and the detunings $\Delta_{\pm}$ are
\begin{align}
|D_{\pm}\rangle=\frac{1}{\sqrt{2}}(|S\rangle\pm|A\rangle),~~\Delta_{\pm}=\Delta\pm\Omega_{SA}.
\end{align}
The Hamiltonian~(\ref{HMR}) in the dressed state basis illustrates that
the original four-level double-$\Delta$ model is simplified to two effective two-level sub-systems, where
the dynamics of each initial chiral ground state is constrained in a two-dimensional
subspace as shown in Fig.~\ref{Fig2}(a) and governed, respectively, by effective two-level Hamiltonian
\begin{subequations}\label{SCIE}
\begin{align}
\hat{H}^{\mathrm{I}}_{L}=&(\sqrt{2}\Omega|L\rangle\langle D_{-}|+\mathrm{H.c.})+\Delta_{-}|D_{-}\rangle\langle D_{-}|,\\
\hat{H}^{\mathrm{I}}_{R}=&(\sqrt{2}\Omega|R\rangle\langle D_{+}|+\mathrm{H.c.})+\Delta_{+}|D_{+}\rangle\langle D_{+}|.
\end{align}
\end{subequations}

For the case with $\phi=0$,
the original four-level
double-$\Delta$ models are simplified to two effective two-level sub-systems
as shown in Fig.~\ref{Fig2}(b) with
\begin{align}\label{HMR2}
\hat{H}^{\mathrm{II}}=&\sqrt{2}\Omega(|L\rangle\langle D_{+}|+|R\rangle\langle D_{-}|+\mathrm{H.c.})\nonumber\\
&+\sum_{\lambda=\pm}\Delta_{\lambda}|D_{\lambda}\rangle\langle D_{\lambda}|.
\end{align}
Correspondingly, the effective two-level models for the two chiral ground states $|L\rangle$ and $|R\rangle$ are
\begin{subequations}
\begin{align}\label{THIIa}
\hat{H}^{\mathrm{II}}_{L}=&(\sqrt{2}\Omega|L\rangle\langle D_{+}|+\mathrm{H.c.})+\Delta_{+}|D_{+}\rangle\langle D_{+}|,\\
\label{THIIb}
\hat{H}^{\mathrm{II}}_{R}=&(\sqrt{2}\Omega|R\rangle\langle D_{-}|+\mathrm{H.c.})+\Delta_{-}|D_{-}\rangle\langle D_{-}|.
\end{align}
\end{subequations}

\section{improved laser-distillation method}
In our recent work~\cite{arXiv.xxx}, we have proposed a very different method with only three coherent operations by using the simplified four-level models. Here, we use the simplified four-level model to improve the laser-distillation method~\cite{PRL.84.1669,PRA.65.015401,JCP.115.5349,JPB.37.2811} and discuss some specific issues which are not involved in the original works~\cite{PRL.84.1669,PRA.65.015401,JCP.115.5349,JPB.37.2811}.
The physical mechanism of our
improved laser-distillation method to achieve the complete
enantio-conversion in equilibrium is as follows.
When the desired chiral ground state evolves back to itself in the excitation step, part of the population
in the undesired chiral ground state is transferred to its corresponding up-level. After this, in the relaxation step with all electromagnetic fields turned off, one half of the population in the up-level will relax to the desired chiral ground state and the other half will relax back to the undesired one. By repeating the pair of steps, the molecules of the desired chirality will accumulate until the enantiopure sample is reached.

Specifically, we can choose the simplified four-level models of Fig.~\ref{Fig2}(a) with $\phi=\pi$ to explicitly show
our improved laser-distillation method for converting the initial chiral mixture
to an enantiopure sample of the desired left-handed chirality in equilibrium.
Each molecule in the initial
chiral mixture is described by the density operator
\begin{align}\label{IS}
\hat{\rho}_{0}=P_{L}|L\rangle\langle L|+P_{R}|R\rangle\langle R|
\end{align}
with $P_{L}+P_{R}=1$. The two chiral ground states evolve following their corresponding Hamiltonian as shown in Eq.~(\ref{SCIE}).
For simplicity, we assume all parameters in each excitation step is time-independent. Then, in each
excitation step, the two chiral ground states experience Rabi oscillations with
different frequencies~\cite{PRA.100.043403}.

{In the case that the left-handed ground state $|L\rangle$ is desired,} the operation time of each excitation step is well designed so that the state $|L\rangle$ experiences integer periods of its corresponding Rabi oscillation.
Since the two chiral ground states have different frequencies of Rabi oscillations, the initial state $|R\rangle$ evolves to
\begin{align}
&|\Psi_{RD_{+}}\rangle=c_{R}|R\rangle+c_{+}|D_{+}\rangle,~~~|c_{R}|^2+|c_{+}|^2=1.
\end{align}
This means part of the population in the initial state $|R\rangle$ is transferred to its corresponding up-level $|D_{+}\rangle$.
Then, the finial
state of each molecule becomes
\begin{align}
\hat{\rho}^{\prime}_{1}=P_{L}|L\rangle\langle L|+P_{R}|\Psi_{RD_{+}}\rangle\langle \Psi_{RD_{+}}|.
\end{align}

In the relaxation step, all electromagnetic fields are tuned off. The molecules in the up-level $|D_{+}\rangle$ equally relax back to the two degenerated chiral ground
states~\cite{PRL.84.1669,PRA.65.015401,JCP.115.5349,JPB.37.2811}.
Then, one half of the population in the up-level $|D_{+}\rangle$
is converted to the desired chiral ground state $|L\rangle$. The other half returns to
the undesired one $|R\rangle$.
The finial state of each molecule after applying one pair of steps to the initial chiral mixture
becomes
\begin{align}\label{Rhore}
\hat{\rho}_{1}=(P_{L}+P_{R}\frac{|c_{+}|^2}{2})|L\rangle\langle L|+P_{R}(1-\frac{|c_{+}|^2}{2})|R\rangle\langle R|.
\end{align}

Repeating $n$ times of the pair of steps, we have
\begin{align}\label{NP}
\hat{\rho}_{n}&=[1-P_{R}(1-\frac{|c_{+}|^2}{2})^{n}]|L\rangle\langle L|\nonumber\\
&+P_{R}(1-\frac{|c_{+}|^2}{2})^{n}|R\rangle\langle R|.
\end{align}
This indicates that molecules in the undesired chiral ground state $|R\rangle$ are gradually
converted to the desired chiral ground state $|L\rangle$ since $|c_{+}|\le 1$. When the
equilibrium is reached  ($n\rightarrow\infty$), all molecules are in the left-handed ground state with
$\hat{\rho}_{n}=|L\rangle\langle L|$.


\begin{figure}[h]
  \centering
  \includegraphics[width=0.9\columnwidth]{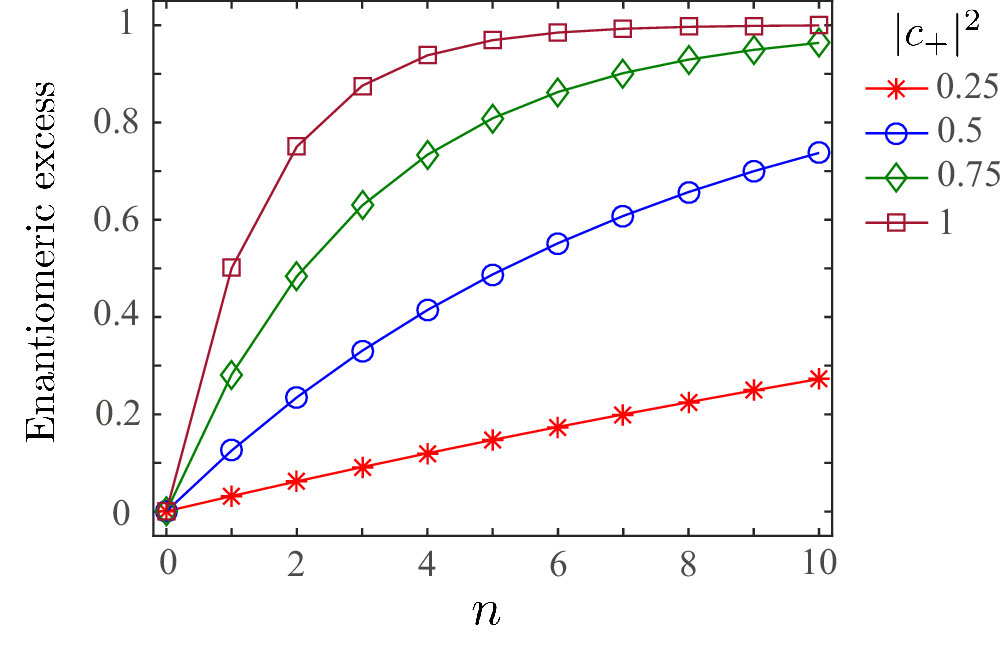}\\
  \caption{Enantiomeric excess with respect to the repeated times $n$ at different $|c_{+}|^2$.
 }\label{Fig3}
\end{figure}
{In realistic case, the repeated times of the two steps should be finite. In Fig.~\ref{Fig3}, we show the enantiomeric excess
\begin{align}
\varepsilon_{n}\equiv\left|\frac{\langle L|\hat{\rho}_{n}|L\rangle-\langle R|\hat{\rho}_{n}|R\rangle}{\langle L|\hat{\rho}_{n}|L\rangle+\langle R|\hat{\rho}_{n}|R\rangle}\right|
\end{align}
after repeating finite $n$ times for different $|c_{+}|^2$.
The results show the tendency of complete enantio-conversion with infinite repeated times. For the finite repeated times (e.g. $n\le10$), we find that the achieved enantiomeric excess is greatly boosted with the increase of $|c_{+}|^2$. In the optimal case with $|c_{+}|^2=1$, after 10 repeated times of the pair of steps, the achieved enantiomeric excess is about $99.90\%$. This guarantees the ability of our improved laser-distillation method for enantio-conversion in the realistic case where the repeated times should be finite.}

\section{Optimization of the excitation step}
As we have pointed out, when the excitation step in our improved laser-distillation method is optimized to ensure that one chiral ground state returns to itself and in the meanwhile
the other chiral ground state is transferred to its corresponding
up-level, the achieved enantiomeric excess after finite repeated times of the pair of steps can be
optimized. In the following, we will pave a way to realize such an optimization of the excitation step.

\subsection{square pulse case}
For the square pulse case where the parameters
are time-independent in the excitation step, each initial chiral ground state
experiences Rabi oscillation with chirality-dependent frequency due to the chirality-dependent of
its corresponding two-level model.
When the excitation step works in the simplified four-level model similar to Fig.~\ref{Fig2}(a) with $\phi=\pi$,
the optimization is achieved under condition~\cite{PRA.100.043403}
\begin{align}\label{CDS}
&\Omega_{SA}=-\Delta=\frac{\sqrt{8n_{L}^2-2(2n_{R}+1)^2}}{2n_{R}+{1}}\Omega>0
\end{align}
with integer $n_{L}>n_{R}\ge0$.
At the end of each excitation step, the initial state $|L\rangle$ evolves back to itself
by experiencing integer $n_{L}$ periods of its corresponding Rabi oscillation. In the meanwhile,
the initial state $|R\rangle$ evolves to $|D_{+}\rangle$ with $|c_{+}|^2=1$ by experiencing
half-integer $(n_{R}+1/2)$ periods of its corresponding Rabi oscillation.

The optimization of the excitation step can also be achieved approximately under the conditions $\Delta_{-}\ll\Omega$ and $\Delta_{+}=0$ by further tuning the fields. Then, the desired chiral ground state
$|L\rangle$ is approximately undisturbed. When the undesired chiral ground state $|R\rangle$ can be
transferred to $|D_+\rangle$ after half-integer periods of its corresponding Rabi oscillation, we approximately achieve the optimization of the excitation step. The key point is under the large-detuning condition $\Delta_{-}\ll\Omega$, the desired chiral ground state is approximately undisturbed and the undesired chiral ground is excited. This can also be used to achieve the optical pumping for enantio-conversion~\cite{yc}, where the fields continuously irradiate the chiral molecules and the high enantiomeric excess can be achieved in the steady state.

\subsection{non-square pulse case}
\begin{figure}[h]
  \centering
  \includegraphics[width=0.9\columnwidth]{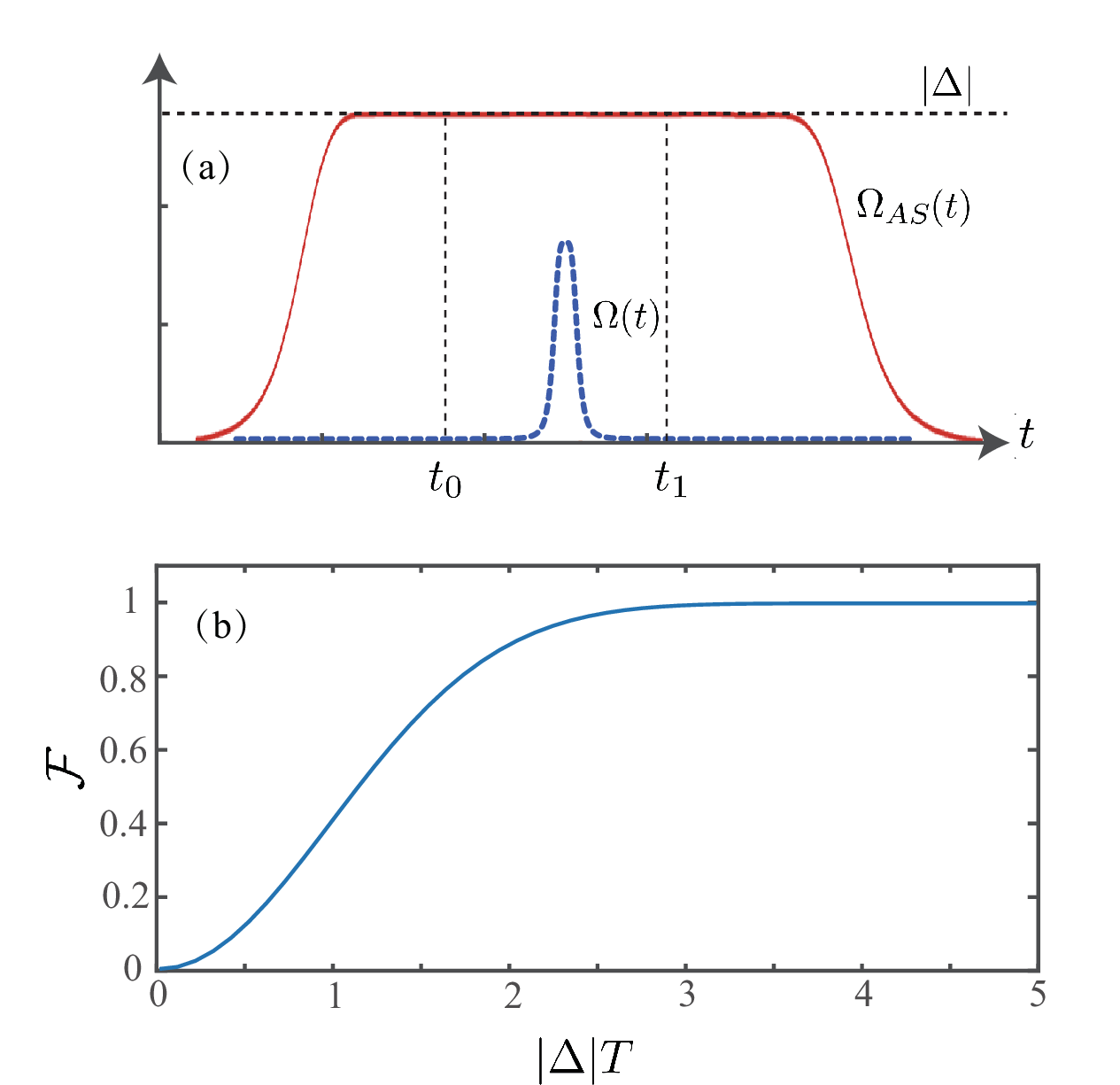}\\
  \caption{(a) Diagrammatic sketch of non-square pulse sequences for the realization of the optimized excitation step within the simplified four-level model similar as shown in Fig.~\ref{Fig2}(a) with $\phi=\pi$. In the time duration of the two electromagnetic fields corresponding to $\Omega~ (=\Omega_{LS}=\Omega_{LA})$, the electromagnetic fields corresponding to $\Omega_{SA}$ should be constant with $\Omega_{SA}=-\Delta>0$. The area of $\Omega(t)$ should be $(2l+1)\pi/2$ with integer $l$, under which
  the state $|R\rangle$ is transferred to $|D_{+}\rangle$ with $|c_{+}|^2=1$.
  (b) Fidelity of our operation under different $\Delta<0$ in the case of $\Omega(t)=({\sqrt{2\pi}}/{4T})e^{-(t-\tau/2)^2/T^2}$ with $\tau=t_{0}+t_{1}$ and $T\ll(t_{1}-t_{0})$. When $\mathcal{F}=1$, the initial state $|L\rangle$ evolves back to itself after our operation and we achieve our optimized excitation step since the area of $\Omega(t)$ is set to $\pi/2$.      }\label{Fig4}
\end{figure}

In realistic experiments,the square pulse is impossible since it would take finite time for the electromagnetic fields to reach the required intensities from the initial zero intensity. In what follows, we will consider the non-square pulse case. In order to realize the optimization of the excitation step, we can first apply the electromagnetic field corresponding to $\Omega_{SA}$ as shown in Fig.~\ref{Fig4}(a). It can be non-square pulse but is constant with $\Omega_{SA}=-\Delta>0$ for a long time.
Then, in the time duration from $t_0$ to $t_1$, we apply the other two electromagnetic fields corresponding to $\Omega~ (=\Omega_{LS}=\Omega_{LA})$. Before the time $t_0$, the chiral mixture is unaffected since the two excited states are unoccupied initially. During the operation from $t=t_0$ to $t=t_1$, within the simplified four-level model similar as shown in Fig.~\ref{Fig2}(a) with $\phi=\pi$, the state $|L\rangle$ evolves in an effective detuned two-level model and the state $|R\rangle$ evolves in an effective on-resonance two-level model.
The evolution matrices of them are
\begin{align}\label{DFU}
&U_{L}\equiv\mathcal{T}\int^{t_1}_{t_0}e^{-i[\sqrt{2}\Omega(t)\sigma^{L}_{x}+\Delta(I^{L}_{0}+\sigma^{L}_{z})]}dt\nonumber\\
&U_{R}\equiv\mathcal{T}\int^{t_1}_{t_0}e^{-i\sqrt{2}\Omega(t)\sigma^{R}_{x}}dt.
\end{align}
{Here, $\sigma^{Q}_{x,y,z}$ are the $2\times2$ pauli matrices and $I^{Q}_{0}$ is the $2\times2$ unit matrix in the corresponding basis of the two chiral ground state $\{|L\rangle,|D_{+}\rangle\}$ and $\{|R\rangle,|D_{-}\rangle\}$.}

In order to realize the optimization of the excitation step, the initial state $|R\rangle$
should be transferred to its corresponding up-level, i.e., $U_{R}=\pm\sigma^{R}_{x}$, which demands
\begin{align}\label{BZD}
\int^{t_1}_{t_0}\sqrt{2}\Omega(t)dt={(2l+1)}\frac{\pi}{2}
\end{align}
with integer $l$. In the large-detuning limit with $|\Delta|\gg\Omega(t)$,
the evolution matrix for the initial state $|L\rangle$ is
$U_{L}\simeq\int^{t_1}_{t_0}e^{-i[\Delta(I^{L}_{0}+\sigma^{L}_{z})]}dt$. Then, the
dynamics of the initial state $|L\rangle$ is approximately frozen. Then, in
each excitation step, the initial state $|L\rangle$ (approximately) keeps unchanged and in the meanwhile the initial state $|R\rangle$ evolves to $|D_{+}\rangle$. With these, the optimization of the excitation
step can be realized.

{In the case that without requiring the large-detuning limit}, the optimization of the excitation
step can be realized by choosing an appropriate $\Delta~(=\Omega_{AS})$ under the condition~(\ref{BZD}).
Specifically, we use Gaussian pulse shape with
\begin{align}\label{GS}
\Omega(t)=\frac{\sqrt{2\pi}}{4T}e^{-(t-\tau/2)^2/T^2},
\end{align}
where we choose $\tau=t_{0}+t_{1}$ and the width of the Gaussian function is $T\ll(t_{1}-t_{0})$.
Under the condition $T\ll(t_{1}-t_{0})$, we have $\int^{t_1}_{t_0}\sqrt{2}\Omega(t)dt\simeq\int^{\infty}_{-\infty}\sqrt{2}\Omega(t)dt=\pi/2$.
Then, we obtain $U_{R}\simeq\sigma^{R}_{x}$ and the initial state $|R\rangle$ can be
transferred to $|D_{+}\rangle$. For the initial state $|L\rangle$, we expect it evolves back to itself. Then, we numerically calculate
the fidelity of our operation
\begin{align}\label{EPI}
\mathcal{F}\equiv|\langle L|{U}_{L}|L\rangle|^2
\end{align}
according to Eq.~(\ref{DFU}) under different $\Delta~(=-\Omega_{AS})<0$ as shown in Fig.~\ref{Fig4}(b). When $\mathcal{F}=1$, the initial state $|L\rangle$ evolves back to itself after our operation and we achieve
our optimized excitation step.
We find with the increase of $\Delta$, the fidelity $\mathcal{F}$ will approach $1$. It is worthy to note that the large-detuning limit is not guaranteed since the peak value of $\Omega$ (about $0.6T^{-1}$) and the value of $\Delta$ for $\mathcal{F}\simeq1$ (e.g. $3T^{-1}$) are both order of $T^{-1}$. Therefore, we have shown that the optimization of the excitation
step can also be realized by choosing an appropriate $\Delta~(=\Omega_{AS})$ in the case that without the requiring of the large-detuning limit.

Until now, our discussions are focused on the situation that the left-handed chiral ground state $|L\rangle$ is desired. {Our discussions can be extended to the situation that the right-handed chiral ground state $|R\rangle$ is desired, since in the framework of our simplified four-level model the dynamics of the two chiral ground states exchanges with each other by changing the overall phase from $\pi$ to $0$ or changing the detuning $\Delta$ to its opposite value.}

\section{Summary}
In summary, we have improved the laser-distillation method~\cite{PRL.84.1669,PRA.65.015401,JCP.115.5349,JPB.37.2811} to realize the complete enantio-conversion when the equilibrium is reached by repeating the pair of excitation and relaxation steps. Our improved laser-distillation method works in the parameter regions, where the original four-level models can be simplified to two effective sub-systems so that the dynamics of each chiral ground state is described by its corresponding two-level model. {With the simplified four-level model, we adopt the optical operations during the excitation step so that the desired chiral ground state evolves back to itself and in the meanwhile part of population in the undesired chiral ground state is excited to the up-level of its corresponding two-level sub-system.
Hence, the laser-distillation method can be improved to realize the complete enantio-conversion in the equilibrium with infinite repeated times of the pair of steps.
By further optimizing the effective detunings of the two effective two-level sub-systems,
the achieved enantiomeric excess after about $10$ repeated times of the pair of steps
can be theoretically increased to about $99.90\%$. We have also paved way for such optimizations in
square pulse case and non-square pulse case.  }

\section*{Acknowledgement}
This work was supported by the National Key R\&D Program of China grant (2016YFA0301200), the Natural Science Foundation of China (under Grants No.~11774024, No.~12074030, No.~U1930402 and No.~11947206), and the Science Challenge Project (under Grant No.~TZ2018003).

{}

\end{document}